\documentclass{aa}  
\usepackage{graphicx}
\usepackage{txfonts}
\usepackage[colorlinks=true,linkcolor=blue, citecolor=blue]{hyperref}
\usepackage{lscape}
\usepackage{amssymb}
\usepackage{amsmath}
\usepackage{soul}
\usepackage[normalem]{ulem}

\newcommand{\msun}{{\rm M_\odot}}
\newcommand{\mjup}{\rm M_{Jup}}

\begin{document}

   \title{The influence of planetary engulfment on stellar rotation in metal-poor main-sequence stars}

   \author{A. Oetjens
          \inst{1,2}
          \and
          L. Carone\inst{1}
          \and
          M. Bergemann\inst{1}
          \and
          A. Serenelli \inst{3,4}
          }

   \institute{ 
              Max-Planck-Institute for Astronomy,
              Königstuhl 17, 69117 Heidelberg\\
              \email{oetjens@mpia-hd.mpg.de}
              \and
              Ruprecht-Karls-Universität, Grabengasse 1, 69117 Heidelberg, Germany
              \and
              Institute of Space Sciences (ICE, CSIC) Campus UAB, Carrer de Can Magrans, s/n, 08193, Bellaterra, Spain
              \and Institut d'Estudis Espacials de Catalunya (IEEC), C/Gran Capita, 2-4, 08034, Barcelona, Spain}

   \date{\today}

\abstract
%Context
{The method of gyrochronology relates the age of its star to its rotation period. However, recent evidence of deviations from gyrochronology relations was reported in the literature.}
% aims
{We study the influence of tidal interaction between a star and its companion on the rotation velocity of the star, in order to explain peculiar stellar rotation velocities.}
% methods
{The interaction of a star and its planet is followed using a comprehensive numerical framework that combines tidal friction, magnetic braking, planet migration, and detailed stellar evolution models from the GARSTEC grid. We focus on close-in companions from $1$ to $20\,\mjup$ orbiting low-mass, $0.8$ and $1\,\msun$, main-sequence stars with a broad metallicity range from [Fe/H] $= -1$ to solar.}
% results
{Our simulations suggest that the dynamical interaction between a star and its companion can have different outcomes, which depend on the initial semi-major axis and the mass of the planet, as well as the mass and metallicity of its host star. In most cases, especially in the case of planet engulfment, we find a catastrophic increase in stellar rotation velocity from $1$ kms$^{-1}$ to over $40$ kms$^{-1}$, while the star is still on the main-sequence. The main prediction of our model is that low-mass main-sequence stars with abnormal rotation velocities should be more common at low-metallicity, as lower [Fe/H] favours faster planet engulfment, provided occurrence rate of close in massive planets is similar at all metallicities.}
% conclusions
{Our scenario explains peculiar rotation velocities of low-mass main-sequence stars by the tidal interaction between the star and its companion. Current observational samples are too small and incomplete, and thus do not allow us to test our model.}
\maketitle
\section{Introduction}

The past decade has brought revolutionary advances in space-based astronomical instrumentation. Space missions, such as Kepler, CoRoT and TESS, enabled astronomers to collect long time series of stellar surface brightness observations. These surveys have provided an unprecedented view on population statistics and physical properties of exoplanet \citep[e.g.][]{Buchhave2012,Fressin2013,Weiss2018,Zhou2019}. They also triggered a massive application of asteroseismology techniques to the photometric data with the aim to constrain the interior structure of stars \citep[e.g.][]{Degroote2010,Beck2012,Chaplin2013,Chaplin2014,Metcalfe2014,Silva2015}.

Combination of data and models enables studies of a coupled physical and dynamical evolution of a planet and its host star \citep[e.g.][]{VanEylen2018,Beck2018}. The star-planet interaction has two major consequences for the structure of the star: the change of the surface chemical composition \citep[e.g.][]{melendezbergemann,Spina2015,Petrovich2017,Veras2019} and the change of the stellar rotation period \citep[e.g.]{Carone2007,Cassan2012,Fleming2019}. Both effects are still poorly-explored with models, and, in particular, it is still not well understood whether and how the angular momentum of a star changes under the influence of the companion. This problem is in the focus of our work.
 
The rotation of a single main-sequence (hereafter, MS) star has canonically been assumed to follow the empirical \citep{skumanich} relation, which relates the rotation rate of a star $\Omega$ to its age, $ \Omega \sim t^{-0.5}$. This relationship was applied to stars in open clusters \citep[e.g.][]{Schatzman1962,bouvier,Barnes2003,Meibom2015} and it forms the basis of standard gyrochronology relationships \citep[e.g.][]{Barnes2007,Angus2015,Claytor2019,Angus2019}. In this model, angular momentum is removed by magnetised stellar winds - the phenomenon known as magnetic braking \citep[e.g.][]{Parker1958,Weber1967,bouvier,Barnes2003}. 

However, recent studies of ages derived by different methods raised controversy over the universal applicability of the Skumanich law. Deviations from this law were observed for a fraction of normal MS stars \citep[e.g.][]{nielsen,sahlholdt}. This can be accommodated within two scenarios: the change in stellar magnetic field configuration \citep[e.g.][]{vanSaders2016,Metcalfe2019} or angular momentum transfer via tidal friction from a close companion \citep[e.g][]{Fleming2019,Santos2019}. Recently, \citet{Amard2020} applied the magnetic braking formalism of \citet{Matt2015} to metal-poor stars, which takes the Rossby number (and hence the convective turnover timescale)  into account. The authors predict deviations from the gyrochronology picture for stars (in a mass range of $1$ - $1.5 \,\msun$) of lower metallicity ([Fe/H] $\lesssim \, -0.5$) due to a lower efficiency in magnetic braking.

In this paper, we focus on changes of stellar rotation\footnote{We note that in the tidal interaction literature it is also common to speak about ``spin'' instead of rotation. } caused by the tidal interaction of a star with a close planetary or low-mass brown dwarf companion using methods and models by \citet{rasio}, \citet{bouvier}, and \citet{Carone2007} and combining them with state-of-the-art stellar structure models \citep{Serenelli2013,Serenelli2017}. We quantify the spin-up rate as a function of the mass and metallicity of a star, as well as the orbit and mass of the companion 

The paper is organised as follows. The model we use to calculate the dynamical interaction between a star and its planet is presented in Sect. \ref{sec:Star_planet_model}. The results of our modelling are discussed in Sect. \ref{sec_results}. We discuss the implications of our results in Sect. \ref{sec_discussion}, summarise our results and present further possible avenues of investigation in Sect. \ref{sec_conclusion}.
\begin{table}[h!] 
\begin{minipage}{\linewidth}
\renewcommand{\footnoterule}{}
\centering
\caption{The notations and variables adopted in this work.}
\label{tab_notation}
\begin{tabular}{lcc}
\hline
\noalign{\smallskip}\hline\noalign{\smallskip} Parameter & Notation & Unit \\
\noalign{\smallskip}\hline\noalign{\smallskip}
\textbf{Star} & & \\

Rotation period & $P_{\star}$ & days \\
Angular frequency of rotation & $\Omega_{\star} = \dfrac{2 \pi}{P_{\star}}$ & rad/s \\
Rotation velocity & $\upsilon_{rot} = \Omega_{\star} R$ & km/s \\
Mass & $M_{\star}$ & g\\
Mass of CE${}^a$ & $M_{env}$ & g\\
Radius at the bottom of CE${}^a$  & $R_{env}$ & cm\\
\noalign{\smallskip}\hline\noalign{\smallskip}
\textbf{Planet} & & \\
Orbital period & $P_{pl}$ & days   \\
Semi-major axis of the orbit & $a$ & AU\\
Initial semi-major axis & $a_{0}$ & AU\\
planetary revolution rate & n & 1/s\\
Mass & $M_{pl}$ & g\\
\noalign{\smallskip}\hline\noalign{\smallskip}
${}^a$ CE denotes the convective envelope
\end{tabular}
\label{tab: notation}
\end{minipage}
\end{table}

\section{Dynamical model for star-planet interaction} \label{sec:Star_planet_model}
We use the star-planet interaction formalism established by \citet{Carone2007}, which combines tidal migration with magnetic braking, in order to quantify the impact of a tidally migrating massive companion with $1 - 20\,\mjup$ on the rotation rate of its host star. We mainly refer to the companion as a "massive planet" and not specifically differentiate between massive planets ($< 13\,\mjup$ for solar metallicity) and low-mass brown dwarfs ($ \geq 13\,\mjup$ for solar metallicity)\footnote{http://w.astro.berkeley.edu/~basri/defineplanet/IAU-WGExSP.htm}. We focus on stars in the MS evolutionary phase and apply the model to stars with masses in the range of $0.8$ to $1 \, \msun$ and metallicities from $-1 \leq$ [Fe/H] $\leq 0 $ dex, in order to check the influence of chemical composition on the tidal interaction. Only massive planets on very tight orbits with less than ~0.03 AU can tidally migrate inwards and spin-up their star within the MS lifetime \citep{Carone2007, Jackson2009}. 

\begin{figure*}
    \centering
    \includegraphics[width=120mm]{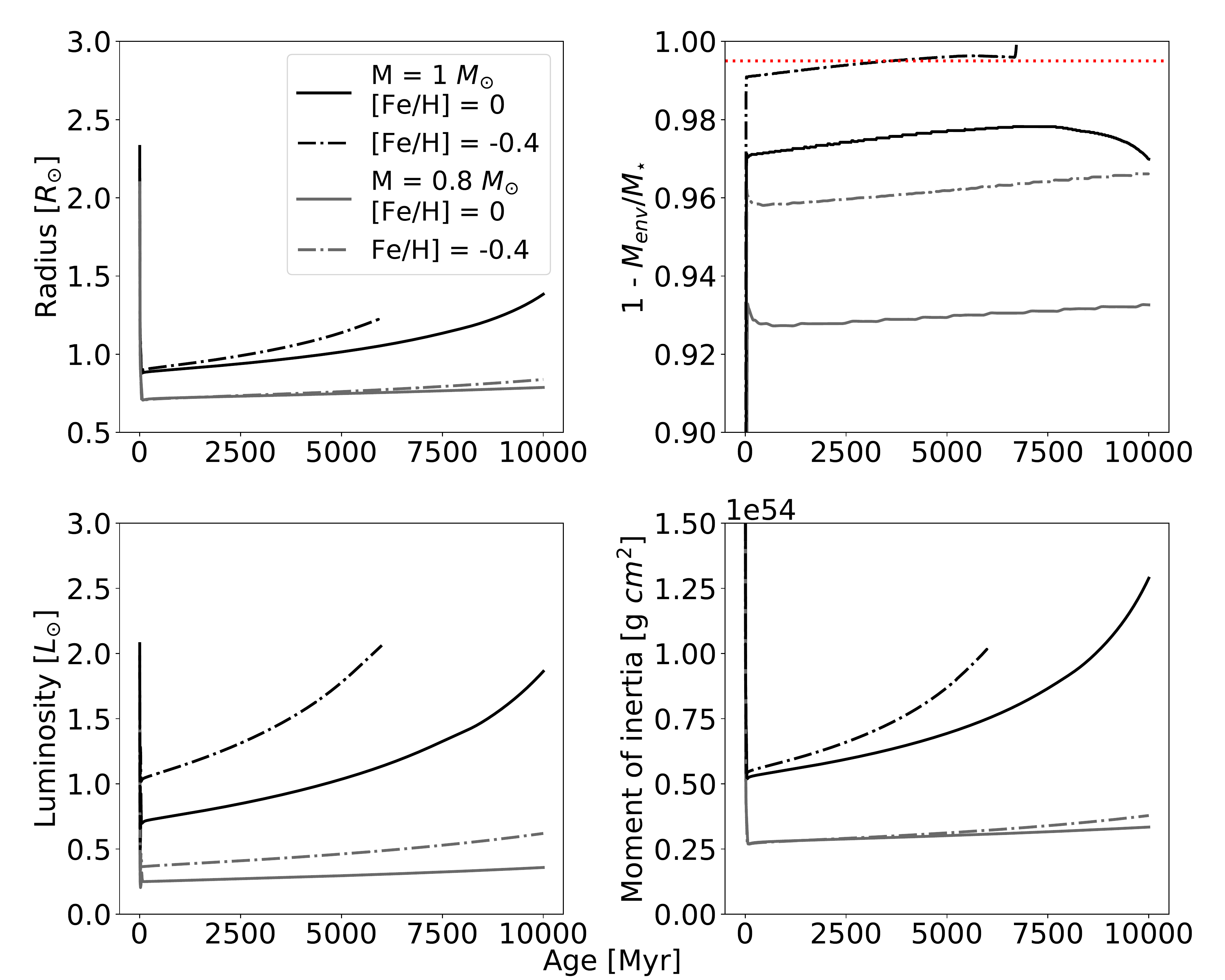}
    \caption{Profiles of different MS stellar parameters against the age of a star for four representative evolutionary tracks computed with the mass of $0.8$ and $1 \, \msun$ and [Fe/H] $=-0.4$ and $0$ (solar metallicity). The top right plot shows the mass of the convective envelope with the limit of $ \rm M_{env} / \msun \, \approx \, 0.005$ indicated by the dotted red line.}
    \label{fig_evolparamMS}
\end{figure*}

All quantities, which describe the global properties of a star as a function of its evolutionary stage -- the mass $M_{\rm env}$ and the radius $R_{\rm env}$ of the convective envelope, stellar radius $R_{\star}$ and luminosity $L_{\star}$ -- are taken from the Garstec stellar evolution models (\citet{Schlattl&Weiss2008}, \citet{Serenelli2013},\citet{Serenelli2017}). The grid of Garstec models covers the full parameter space of low-mass stars: mass in the range from $0.6$ to $5\,\msun$ and metallicity [Fe/H] in the range from $-2.5$ to $+0.6$ in increments of $0.05$ dex.This study focuses on the parameter space of $0.8$ and $1 \, \msun$ and [Fe/H] from $-1$ to $0$. The profiles of different stellar parameters for several evolutionary tracks are shown in Figure~\ref{fig_evolparamMS}.
\subsection{Gain of angular momentum through tidal friction}
\label{sec: tidal friction}
We focus on the effect of the equilibrium tide on the star, which best describes tidal interactions of a companion with a low mass star of $1\,\msun$ and less \citep{Fleming2019}. Furthermore, we use the tidal friction formulation first established for binary stars by \citet{Zahn1977} and expanded by \citet{rasio} and \citet{privitera} to extrasolar planets around MS stars, which allows to self-consistently combine the efficiency of tidal migration with stellar structure. In this formulation, tidal dissipation occurs predominantly in the convective envelope of a star. The frictional loss of energy is then extracted from the orbit of the companion with the semi-major axis $a$ and planetary revolution rate $n$. We can safely assume a circular orbit because $e\approx 0$ is quickly achieved via efficient eccentricity damping for planet-mass objects on orbits with semi-major axis of 0.03~AU and shorter \citep{Dobbs2004}\footnote{See also exoplanet.eu that shows $e\approx 0$ for the vast majority of hot Jupiters on tight orbits .}. The rate of orbital decay is given by:
\begin{equation}
 \left( \frac{\dot{a}}{a}\right)_{\rm tidal}  =  \frac{f}{\tau} \, \frac{M_{\rm env}}{M_{\star}} \, q \, \left( 1 \, + \, q \right) \, \left( \frac{R_{\star}}{a}\right) ^{8}, 
 \label{eq: dot_a} 
\end{equation}
where $\tau$ is the convective turnover time scale, $M_{\rm env}$ the mass of the convective envelope of a star at a given point in time, $M_{\star}$ and $R_{\star}$ the total mass and radius of a star, and q the planet-to-star mass ratio $q = M_{\rm pl}/M_\star$. 

The numerical factor $f$ is given by 
\begin{equation}
f = \min\left[1,\left(\frac{P_{pl}}{2\tau}\right)^2\right]
\end{equation}
and it quantifies viscous dissipation of tidal energy throughout the convective zone. $P_{\rm pl}/2$ is the so-called tidal pumping frequency and accounts for the occurrence of two tidal bulges during one orbital revolution of the planetary companion around its star.  If the convective turnover timescale $\tau$ is short compared to the tidal pumping frequency, dissipation is very efficient and $f=1$. Otherwise, only those eddies with timescales shorter than $P_{pl}/2$ can contribute to the dissipation of energy and $f$ is smaller than 1 \citep{Goldreich1977,rasio}.

The convective eddy turnover timescale is calculated using the equation suitable for large eddies at the base of the convective zone \citep{rasio,privitera}:
\begin{eqnarray}
\tau \; = \; \left[ \frac{M_{env} \, \left( R_{\star} \, - R_{env} \right) ^2}{3 \, L_{\star}} \right]^{1/3},
\label{tau}
\end{eqnarray}
where $L_{\star}$ is the luminosity of a star. 

We note that in large parts of the tidal interaction literature \citep[see e.g.][]{Jackson2009,Barnes2015,Yee2020} instead of $\tau$, the tidal dissipation factor $Q_{\star}$ or modified tidal dissipation factor  $Q'_{\star}=Q_{\star}/{k_{2,\star}}$ is used to quantify the efficiency of tidal interactions, where the latter comprises also the stellar Love number $k_{2,\star}$. To facilitate comparison with other works, we thus also introduce how to derive the modified tidal dissipation factor $Q'_{\star}$ in the framework of this work via:
\begin{equation} \label{dissipation_factor}
\centering
Q'_{\star} = \frac{3GM_{\star}}{R^{3}_{\star} n} \frac{\tau}{f} \frac{M_{\star}}{M_{env}}
\end{equation}

For comparison with other tidal interaction works, we further calculated the modified stellar tidal dissipation factor and derived $ Q_\star' = 10^{7}$ for the systems that we studied. This $Q_\star'$ values fits well into the range of $Q'$ values used in the literature  \citep[see e.g.][]{Barnes2015, Heller2019}.
 
\begin{figure}[hb!]
\centering
\includegraphics[width=0.8\columnwidth]{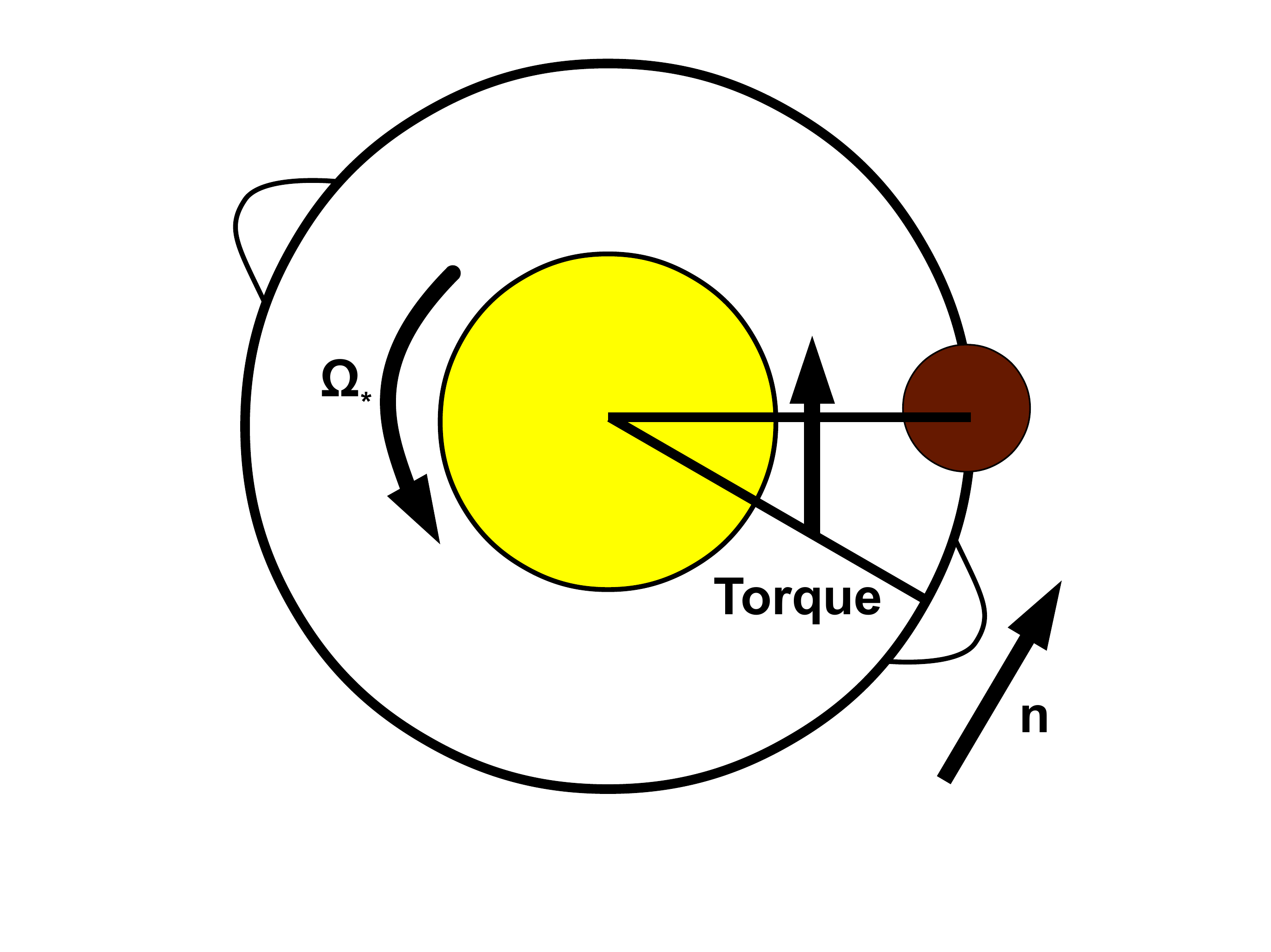}
\caption{Angular momentum transfer during planetary engulfment due to tidal friction. See also Figure~1 in \citet{Carone2007}.}
\label{fig_tidal_spin_up}
\end{figure}

In tidal theory, many equations, like the differential equation for the semi-major axis evolution description used in \citet{Carone2007}, compare the orbital revolution rate $n$ with the stellar rotation rate $\Omega_\star$. The revolution rate $n$ is more informative in the framework of revolving and rotating bodies of unequal mass, where two bodies (one planetary mass and one stellar mass) orbit around the common centre of mass. The centre of mass lies however, inside the host star such that we can assume, to first order, that the the planet revolves around the host star (see Figure~\ref{fig_tidal_spin_up}).

In this framework, orbital decay and stellar spin-up only occur if the stellar rotation rate is lower than the planetary orbital revolution rate $n$. Or in other words, the orbit decays if the stellar rotation period is larger than the orbital period of three days and less for companions on very tight orbits as considered in this work. If the stellar rotation rate and revolution rate are equal, that is, synchronized, tidal friction and thus orbital decay stops in the absence of magnetic braking.

Given the tidal evolution of the semi-major axis (eq. \ref{eq: dot_a}), we can derive the equations for the rate of change of the stellar rotation rate $\Omega_{\star}$  via angular momentum transport of the orbit towards the host star. The dominant fraction of the planetary angular momentum is stored in the planetary orbit. Therefore, changes in the planetary spin can be neglected \citep[e.g][]{carone, Murray}. Thus, the total angular momentum $L$ of the star - planet system mainly consists of the component stored in the stellar spin and in the companion orbit \citep[cf.][]{Murray}:
\begin{equation}
L= I_\star \Omega_\star + \frac{M_\star M_{pl}}{M_\star + M_{pl}} a^2 n \label{eq: tot_ang_mom},
\end{equation}
where $I_{\star}$ is the moment of inertia of the star defined as $I=k^{2}M_\star R_\star^{2}$ and $n$ is the orbital revolution rate again.

When the equations~\ref{eq: dot_a} and \ref{eq: tot_ang_mom} are combined\footnote{Here, we assume that $\dot{L}=0$ and $\dot{I}\approx  0$ during the MS. Further, it can be shown that $\frac{d}{dt}(a^2n) = \frac{1}{2} a \dot{a} n$, using Kepler's law $n=a^{-3/2}\sqrt{G(M_{Pl}+M_\star)}$},  the tidal spin-up can be approximately described by:
\begin{eqnarray}
\left( \frac{d\Omega_\star}{dt} \right)_{tide} & = & - \frac{M_{pl}M_\star}{2 I_\star (M_{pl}+M_\star)} n~a~\dot{a}   \\
& = & - \frac{1}{2 \, k^2} \, \frac{P_{pl}}{2 \, \tau^3} \, \left(\frac{M_{pl}}{M_{\star}}\right)^2 \, \left( \frac{R_{\star}}{a} \right) ^6 \, \frac{M_{env}}{M_{\star}}  \label{eq: omega_tide},
\end{eqnarray}
The $k^{2}$-value ranges from 0.03 for fully radiative stars to 0.2 for fully convective stars \citep{nelemans}. For all calculations where we focus on MS stars with a thick enough convective outer envelope that promotes efficient tidal interaction, $k^{2}$ is set to the solar value of 0.074, which was calculated based on moment of inertia calculated from the mass-radius distribution based on the helioseismology data of \citet{Dziemnbowski1994}. We confirmed in our stellar models that $k^{2}$ differs at most by 30\% from this value. See also \citet{bouvier} their Figure~1, the authors of which also focus on MS stars with masses of $0.8$ and $1 \, \msun$. 
          
The tidal model adopted in this work is valid for stars with a thick convective envelope that allows for efficient tidal dissipation of energy. The efficiency limit of stellar tides can also be inferred from their effects on the obliquity in a planetary system, that is the angle between a planetary orbital plane and the stellar rotation axis. The obliquity can be measured via the Rossiter-McLaughlin effect during the transit of close-in giant planets (or hot Jupiters). See \citet{Queloz2010} for the first such measurement for an exoplanet. \citet{Winn2010, Albrecht2012} pointed out that based on the measured obliquities of Jupiters, it appears that originally shortly after formation the obliquities of hot Jupiters can be very large. These large obliquities are then damped afterwards to small values via tides, provided the host star is cool enough to have a thick convective envelope. The limit of tidal efficiency was identified to be at $M_{\rm env}/M_{\star} \approx 0.005$ \citep[][Fig.2]{Winn2010}, and for smaller values tides are negligible. For stars of solar metallicity and higher this corresponds to strong tides and small obliquities for $T_{\rm eff} < 6250$~K, in accordance with observations of the stellar spin-orbital plane angle for systems with hot Jupiters via the Rossiter-McLaughlin effect \citep{Winn2010}. We have taken the mass and temperature thresholds into consideration and for metal-poor stars, which are hotter, exclude those cases for which $M_{\rm env}/M_{\star} \approx 0.005$.
\subsection{Loss of angular momentum by magnetised stellar winds}
\label{sec:rot}
\begin{figure}[h!]   
    \centering
    \includegraphics[width=90mm]{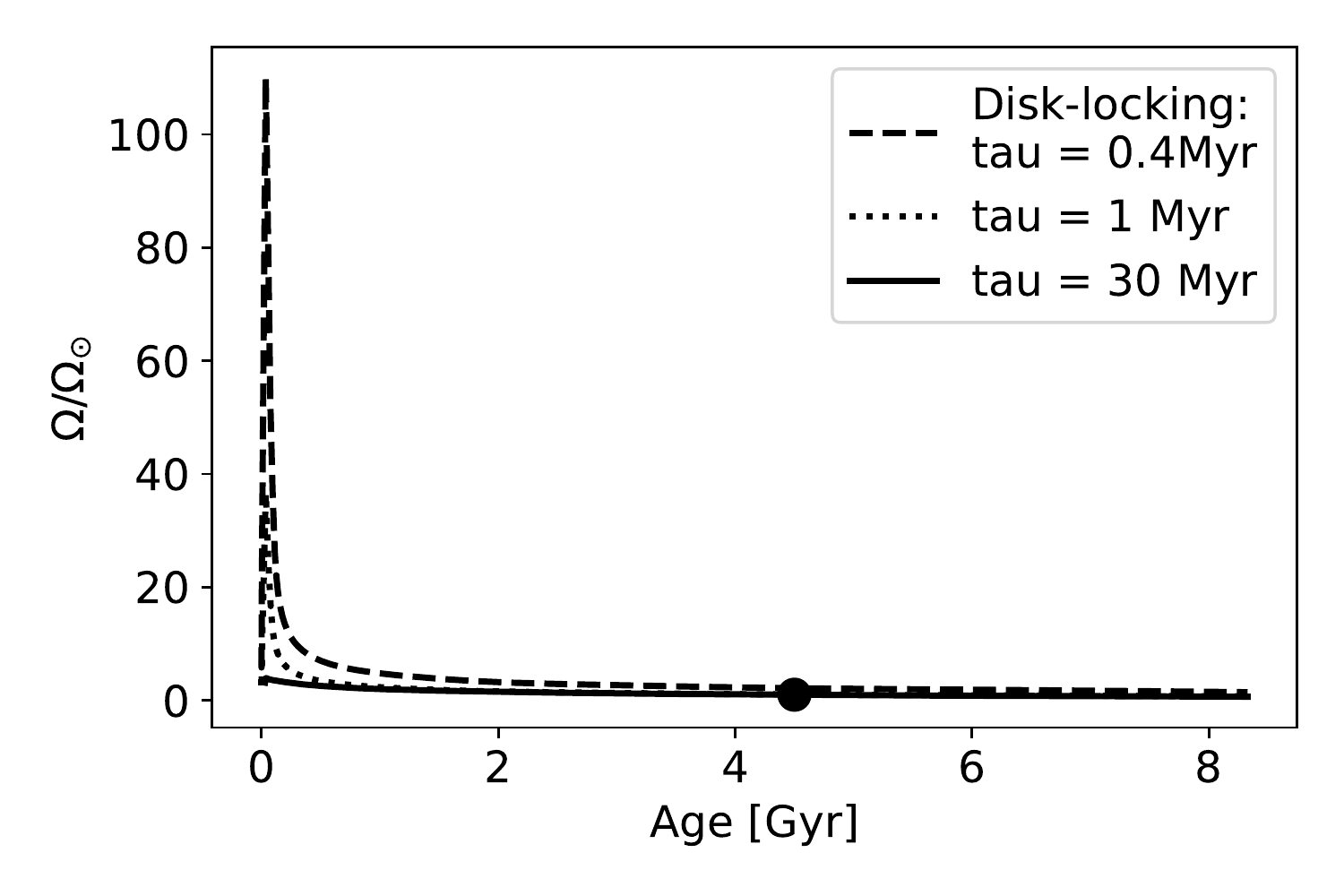}
    \caption{One example simulation of rotation velocity according to \citet{bouvier} for a 1 $\msun$ model. The black dot indicating the solar value at $4.5$ Gyr. The input parameters are presented in table 2. Shown are three different curves for different disk locking time, i.e. the time it takes the star to decouple from its surrounding protoplanetary disk.}
      \label{fig_bouvier}
\end{figure}

To coherently study stellar evolution over the MS, we also take into account the evolution of stellar rotation rate caused by the loss of angular momentum by magnetised winds. In the following, we use the model of \cite{bouvier}:
\begin{eqnarray}
\frac{1}{\Omega_{\star}}\left(\frac{d\Omega_{\star}}{dt} \right)_{wind} & = & \frac{1}{J}\left(\frac{dJ}{dt}\right) \: - \: \frac{1}{I}\frac{dI}{dt},
\label{bouv3} 
\end{eqnarray}
where $J$ is the angular momentum of the star $J  =  I\,\Omega_{\star}$, assuming solid body rotation. Stellar evolution models are used to calculate the evolution of $dI/dt$ with time. The change of angular momentum of a star is, furthermore, described by :
\begin{eqnarray}
\frac{dJ}{dt}  & = & -K\Omega_{\star}^{3} \left(     
\frac{R_{\star}}{R_{\odot}}  \right) ^{1/2} \left( \frac{M_{\star}}{\msun}\right)^{-1/2} \label{bouv1} \quad \quad \;    \text{for}\: \Omega_{\star} \; \leq \;  \omega_{sat}  \\
\frac{dJ}{dt}  & = & -K\omega_{sat}^{2}\Omega_{\star} \left(          \frac{R_{\star}}{R_{\odot}} \right) ^{1/2} \left( \frac{M_{\star}}{\msun}\right)^{-1/2} \label{bouv2}           \quad \text{for} \: \Omega_{\star} \; > \;  \omega_{sat} 
\end{eqnarray}
where $K$ is a phenomenological scaling factor that \citet{bouvier} introduced to connect  observational rotation rates from stellar clusters of different ages to the observed small rotation rates of MS stars like the Sun.  $\omega_{sat}$ is the rotation rate, for which the magnetic field is assumed to saturate. With faster rotation, magnetic activity and thus magnetic braking no longer increases strongly, following the dynamo relation at the basis of this formulation. Consequently, also the stellar angular momentum loss saturates and no longer increases with faster rotation. $\omega_{\rm sat}$ is typically constrained by observations to explain the spread in rotational velocities in ZAMS clusters. It was scaled by \citet{bouvier} to be $14 \Omega_\odot = \Omega_{sat}$ for $1\,\msun$ stars. Although the scaling varies with mass, we adopt the before mentioned value to make the computations more comparable to other studies in the field. \citet{Amard2020} adopt a single value as well, $\Omega_{sat} = 11 \Omega_\odot$ for masses between $0.8 - 1.3\,\msun$ stars. The evolution starts with $P_{\rm ini}$. The values for $K$, $\omega_{\rm sat}$ and $P_{\rm ini}$ for different masses are listed in table~\ref{bouvparam}. 
Figure~\ref{fig_bouvier} shows the application of the Bouvier model and that by the time a star reaches solar age, regardless of the initial conditions, the magnetic braking has sufficiently slowed down the star to solar value. 

{}

During the tidal migration, the total effect of magnetic braking and tidal interaction is calculated with:
\begin{eqnarray}
\label{w_spin_up}
\frac{d\Omega_{\star}}{dt} & = & \left( \frac{d\Omega_{\star}}{dt} \right)_{wind} -\, \frac{1}{2 \, k} \, \frac{P_{pl}}{2 \, \tau^3} \, \left(\frac{M_{pl}}{M_{\star}}\right)^2 \, \left( \frac{R_{\star}}{a} \right) ^6 \, \frac{M_{env}}{M_{\star}}
\end{eqnarray}
Although we assume that tidal dissipation takes place mainly in the outer convective envelope of the star, the angular momentum transfer from the planetary orbit spins-up the star as a whole, assuming solid body rotation.
\begin{table}[hb!] 
\begin{minipage}{\linewidth}
\renewcommand{\footnoterule}{}
\centering
 \caption{Parameters for the stellar rotation evolution.}
 \label{bouvparam}
\begin{tabular}{l c}
\hline
\noalign{\smallskip}\hline\noalign{\smallskip} Parameter & Value \\
\noalign{\smallskip}\hline\noalign{\smallskip}
Disk locking time ($\tau_{disc}$) & 1 Myr    \\
Saturation velocity ($\omega_{sat}$) &  14  $\Omega_{\odot} $         \\
Scaling factor (K)      &  2.7 $\times 10^{47}$  \\
Initial rotation period ($P_{ini}$)   & 7.8 days            \\
Initial angular velocity ($\Omega_{0}$) & 9.32 $\times 10^{-6}$ s$^{-1}$ \\
Solar angular velocity ($\Omega_\odot$) & 2.9 $\times 10^{-6}$ s$^{-1}$ \\
\noalign{\smallskip}\hline\noalign{\smallskip}
\end{tabular}
\end{minipage}
\end{table}

We note that there have been recent improvements on the Bouvier model to improve stellar evolution during the pre-main-sequence more coherently \citep{Gallet2013,Gallet2015}. Since we focus here on the effect of tidal migration ($\tau_{tide}\geq 1$~Gyrs), which affects the star only when it is already on the MS, the simpler Bouvier model is sufficient for this work.

We note also that the Bouvier model that we use here is based on large parts on the model of \citet{Kawaler1988}. The latter model is used in recent work by \cite{Amard2020} to represent spin-down rates due to "nominal" magnetic braking compared to reduced magnetic braking for massive, metal-poor stars ($M_{\star} \geq 1\,\msun$) over time scales of billion of years. \citet{Amard2020} also stress the importance of stellar structure in their work, which indicates that the absence of a thick convective envelope is responsible for reduced magnetic braking.

Conversely, in this work, we focus on less massive ($M_{\star} \leq 1 \, \msun$), metal-poor stars, because for efficient tidal interactions a thick enough convective envelope is required. A thick convective envelope promotes at the same time efficient magnetic braking as we simulate in this work. We stress again that we always check with stellar modelling that a convective layer thicker than $0.005 \, M_\star$ is present, when applying our star-planet interaction models.
\subsection{Stellar spin-up due to tidal friction}
\label{sec_spin-up}

As mentioned before, we consider tidal migration of sub-stellar mass companions on tight orbits ($0.03$ AU and shorter or for orbital periods shorter than 3~days). At the same time, the star loses its angular momentum via magnetic braking.
           
The choice of initial conditions is motivated by the following considerations. A Sun-like star with a mass of $0.8 - 1 \, \msun$ and the age of 4.5 Gyr typically rotates with the velocity of $25.1$~days ($\Omega = \Omega_\odot$). The pre-main-sequence phase of a star is much shorter than the characteristic timescales of tidal migration ($\tau_{tide}\geq 1$~Gyr). Therefore, we assume for the tidal migration scenario that the planetary orbital period $P_{\rm pl}$ is initially significantly smaller than the stellar (surface) rotation period $P_{\star}$ at the time when the sub-stellar companion starts to migrate inwards. In such a situation, the crest of the tidal bulge on the star, induced by the companion, always lags behind with respect to the line connecting the centre of masses of the companion, the star, and the tidal bulge. Therefore, the tidal torque acts in the direction of stellar rotation leading to the spin-up of the star. The orbit of the companion shrinks due to the conservation of the total angular momentum (Figure~\ref{fig_tidal_spin_up}).

\begin{figure}[h!]
\centering
\includegraphics[width = 85mm]{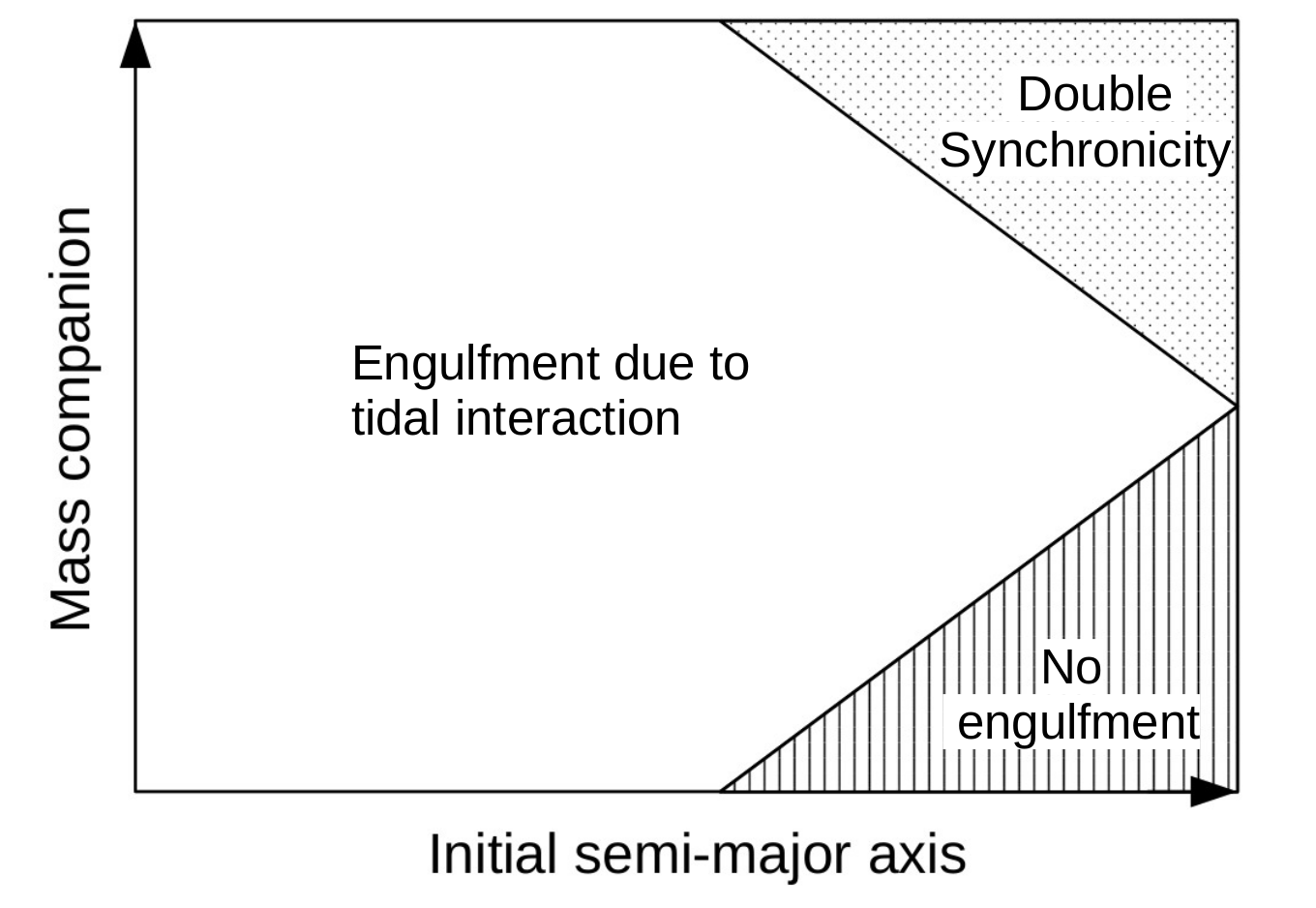}
\caption{The outcomes of tidal interaction between a star and its companion in our model: engulfment (white area), double synchronicity (upper right corner), and no significant interaction up until the MS turnoff point (lower right corner)}
\label{fig_legend}
\end{figure}

The decay of the planetary orbit depends on two quantities. One is the total angular momentum $L$ in the system. The other is the relationship between the orbital revolution rate $n$ of the planet and the rotation rate of the star $\Omega_\star$ (Table \ref{tab_notation}). In this work, we consider two cases: the tidal inwards migration ($\Omega_\star < n$) and the double synchronicity ($\Omega_{\star} = n$). Both scenarios are shown in Figure~\ref{fig_legend} and will be briefly described below. If $\Omega_{\star} > n$, that is the companion is beyond a critical distance, the tidal migration is reversed and the companion migrates outward.  In this configuration, angular momentum is transferred from the host star to the companion and, in fact, a decrease in the stellar rotation rate occurs. A similar phenomenon occurs in the the Earth-Moon system, where tidal friction acts to slow down the Earth's rotation as the Moon migrates away from the Earth.

\subsubsection{Tidal inwards migration}
In this scenario, tidal friction causes the companion (a planet or a brown dwarf) to migrate inwards and to gradually spin-up the star. We run the simulation up until the point when the star has approached the MS turnoff point or when the planet has reached the Roche limit, the distance at which a planet will disintegrate due to the strong differential of tidal forces inside the body exceeding the planetary gravitational self-attraction which is defined as:

\begin{equation}
d_{\rm Roche} = R_{\star}\left(2 \frac{\rho_{\star}}{\rho_{\rm pl}}\right)
^{1/3}
\end{equation}
The Roche limit varies between 1.44 and 2.44 $R_{\star}$ \citep{Murray}. We adopt a conservative assumption that the Roche limit is located at $1.44$ $R_\star$ and that angular momentum is no longer transported towards the star upon the disruption of the planet. That is, we conservatively assume that upon disruption, the remaining angular momentum of the planet is removed from the system. If some angular momentum is still transported to the star during disruption, the star would be spun-up even faster. Therefore, the final rotation rate of a star obtained in our simulations is a lower limit\footnote{As a reminder, higher rotation rates mean faster rotation and shorter rotation periods.}

\begin{figure}[hb!] 
\centering
\includegraphics[width=90mm]{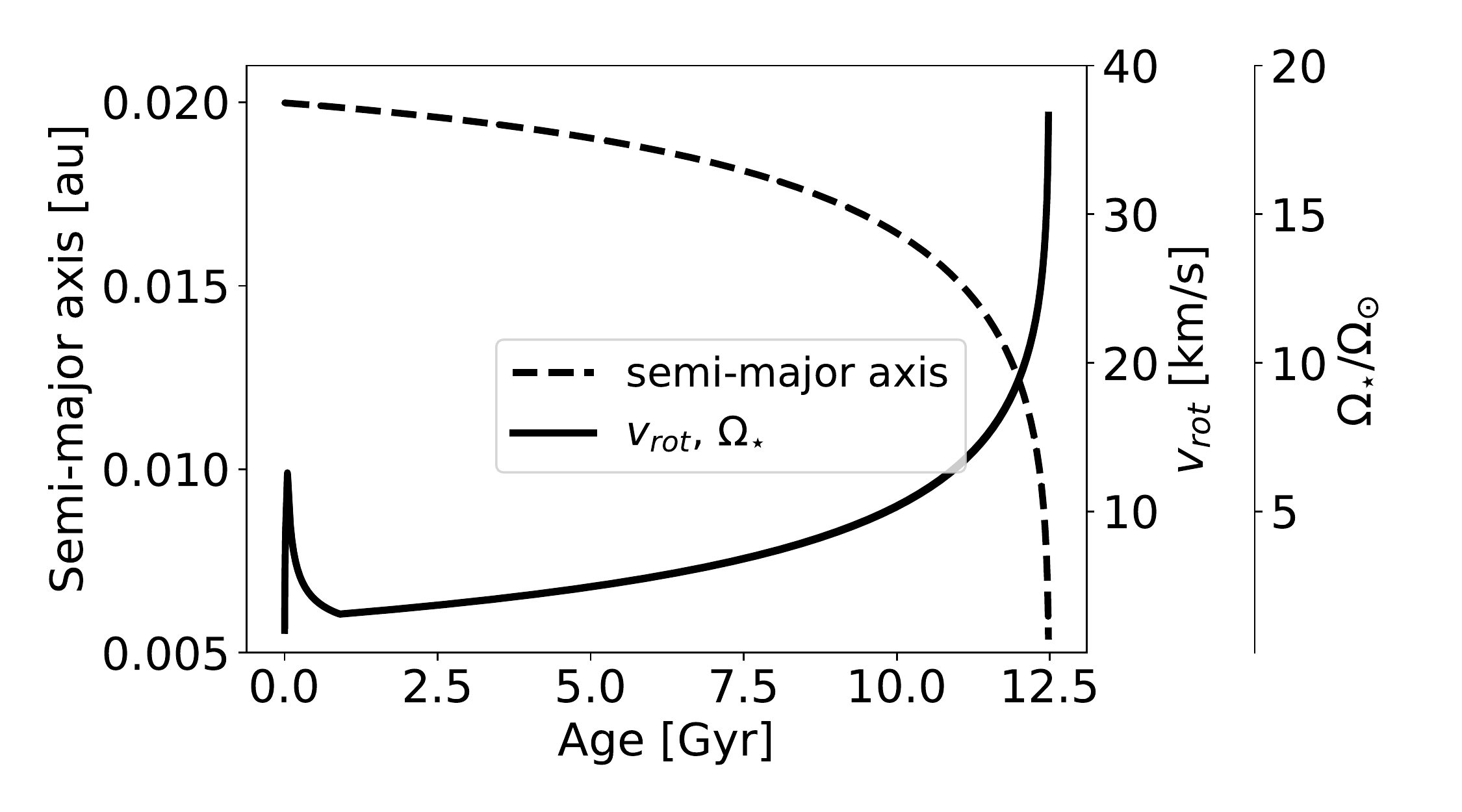}
\includegraphics[width=90mm]{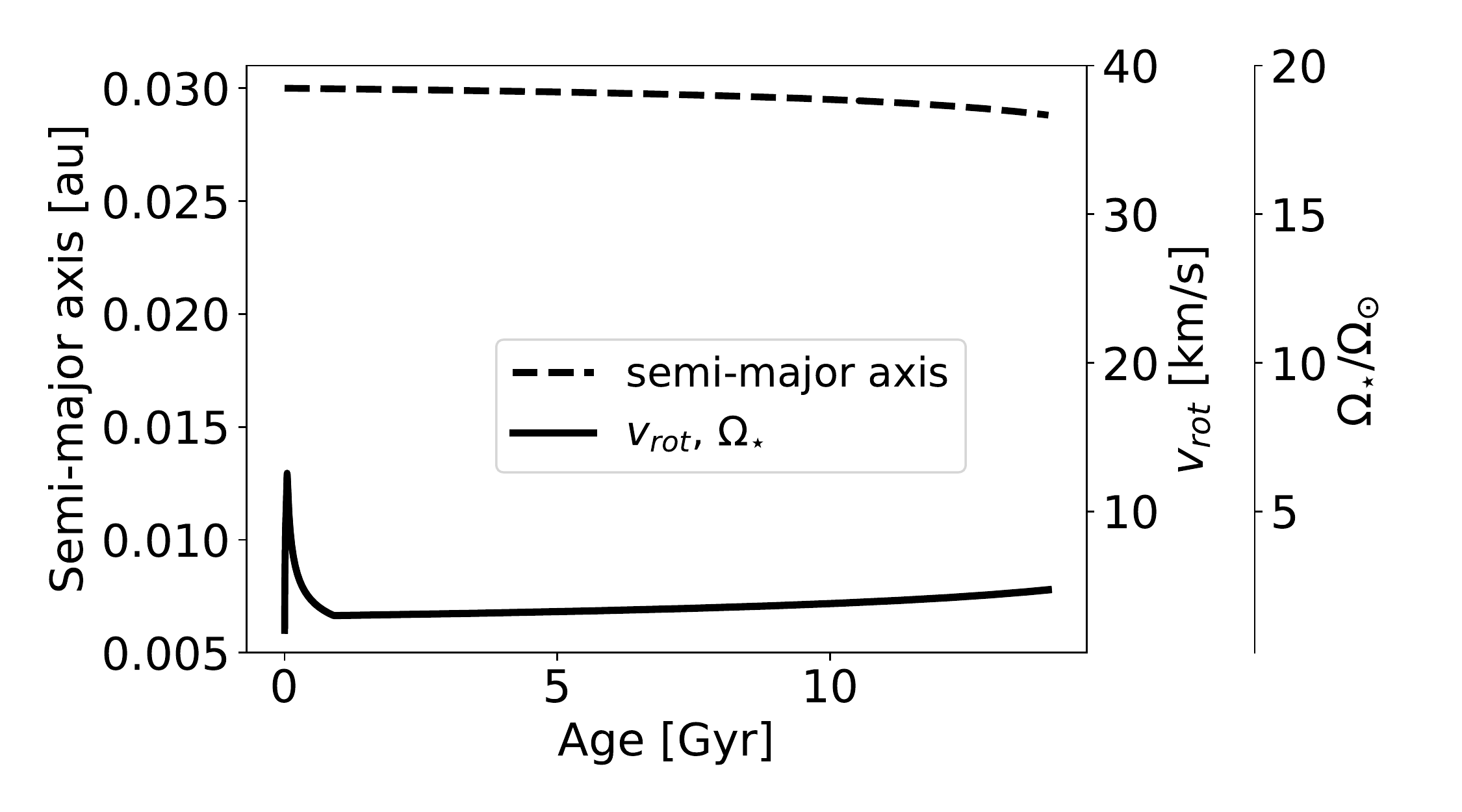}
\caption{Top: evolution of the semi-major axis (dashed line) \ and the rotation velocity (solid line) of a $0.8 \, \msun$ star with [Fe/H] $= -0.4$ and a 10 $\mjup$ companion at a distance of $a_0 = 0.02$ AU. Bottom: Same conditions for the same stellar model with a 10 $\mjup$ companion at $a_0 = 0.03$ AU. No engulfment occurs during the MS lifetime.}
\label{fig_vrot_evol}
\end{figure}

Further, the time scale of planetary migration is one of the most important quantities in our model, because it defines two possible outcomes of the dynamical interaction, which we refer to as the fast and the slow migration. They are illustrated in Figure~\ref{fig_vrot_evol}. If the timescale of tidal migration is shorter than the MS lifetime of a star, $\tau_{\rm tide} \leq \tau_{\rm life}$, the sub-stellar companion reaches the Roche limit within the MS lifetime of a star and is destroyed (Fig.~\ref{fig_vrot_evol}, top panel). On the other hand, if $\tau_{\rm tide}>\tau_{\rm life}$, the sub-stellar companion does not reach the Roche limit within the MS lifetime of the star. In both cases, the rotation velocity of the host star can be significantly increased (Fig.~\ref{fig_vrot_evol}, bottom panel).
Previous work in this field, see e.g. \citet[][Fig.2,Fig.4]{Carone2007} and \citet[][Fig.6c]{Bolmont2017} explore the effects of tidal interaction and magnetic braking in the context of planetary engulfment. The amount of angular momentum stored in the planetary orbit is much larger compared to the stellar rotation, which reduces the efficiency of the magnetic braking mechanism. At the same time, magnetic braking tends to saturate for fast rotating stars (here $14 \Omega_\odot$) and thus can not compensate for the still increasing angular momentum transfer as the companion moves further in. The details of how much the star can be spun-up, however, depend on when planetary engulfment (and thus angular momentum transport) is completed. E.g. \citet{Carone2007} assumed that planetary engulfment stops at a Roche limit of $2.4 R_\star$ for a 1.5 $\mjup$ planet, \citet{Bolmont2017} stop planetary engulfment at 0.01~AU for a 1 $\mjup$ planet. In this work, we assume a Roche limit of $1.44 R_\star$.

\subsubsection{Double synchronicity}
If $\Omega_\star = n$, that is, when the stellar rotation rate is equal to the planetary revolution rate, there is no tidal friction and thus no torque that can spin-up the star or lead to the orbital decay. \citet{Hut1980} showed that when the total angular momentum of a binary system exceeds a critical value $L_{\rm crit}$ the system can end up in a double synchronous rotation and thus, to first order, it is in a stable configuration preventing further migration inwards. We stop the calculations once $\Omega_\star = n$, that is, the double synchronicity is reached for $L>L_{\rm crit}$. It is, in principle, possible that the angular momentum of the system drops below $L_{\rm crit}$ later in the evolution of the star   \citep{carone,Guillot2014}, because of further angular momentum loss due to magnetic braking. The set of tidal equations that we use here are not suitable to simulate the further evolution of a double synchronous state where, a) the tidal bulges raised by the planet no longer move with respect to the surface of the star and b) the dynamical tide and not the equilibrium tide becomes important \citep{ogilvie}. Thus, a separate treatment and different set of tidal equations is required \citep{carone,Guillot2014}. We do not consider this scenario in this work, but postpone its treatment to the next paper in the series. We note, however, that previous work suggests that such systems may only be pseudo-stable and will eventually destabilize, if $L<L_{\rm crit}$ is reached within the life time of the system \citep{Hut1981,carone,Guillot2014}. In this case, the planetary companion will start to migrate inwards again \citep{Damiani2016,Hodvzic2018}.
\subsection{Tidal interaction during the PMS}
\label{sec: PMS}
While we focus on the tidal interaction during the MS, it is worthwhile discussing the impact of planet migration during the PMS and how this may affect the occurrence rate of close-in massive planets and brown dwarfs on the MS.

During the PMS, the equilibrium tide plays a negligible role and the dominant cause for tidal interaction is the dynamical tide. After reaching the MS, the dynamical tides are surpassed in strength by the equilibrium tides, which justifies our decision to neglect dynamical tides in this work. For an in-depth discussion and derivation the reader is referred to \citet{Rao2018, Bolmont2017, Heller2019}. They further suggested that dynamical tides during the PMS are so efficient that many close-in planets (located within 0.04~AU) around solar-mass stars could either be tidally engulfed during the PMS or migrate outwards. In both cases, no planetary engulfment occurs during the MS. In the first case, the planet no longer exists while the star is on the MS and in the second case, the planet is too far away for the weaker equilibrium tides to achieve planetary engulfment during the MS.

As noted by the authors, the exact tidal migration scenario depends very strongly on the initial stellar rotation period. It was consistently shown that mainly planets orbiting stars with initial stellar rotation periods of 2~days and shorter are prone to either get engulfed within the PMS or migrate too far outwards for later planetary engulfment during the MS. Furthermore, dynamical tides in metal-poor stars ([Fe/H]=-0.5) are much less efficient. Thus, even planets on short orbits - within 0.04~AU - could survive while the star is in the PMS evolutionary stage, in particular, when they orbit a host star with an initial stellar rotation period of 6~days and longer \citet[][Fig.7]{Bolmont2017}. In addition, observed stellar rotation in young ($\lesssim$ 5~Myrs) stellar clusters show that a substantial fraction of sun-like stars appear to have rotation periods of 6~days and longer \citep{bouvier,Irwin2008}. Also, while the findings of \cite{Rao2018} and \cite{Bolmont2017} predict the engulfment of companions before reaching the MS for certain conditions, there is evidence that these systems do exist. WASP-19 is a solar like MS-star with a close-in (0.017 AU) 1.2 $\mjup$ companion \citep{Hebb2010}.

In conclusion, tidal interaction between the star in the PMS evolutionary stage and its planetary companion can have a significant effect on individual planetary systems. However, in this work, we want to stress that stellar spin-up during the MS due to tidal migration on the MS can occur and could be observable due to excess of rotational velocity in the host star. The observation of abnormally fast (exceeding model predictions) rotators on the MS would then justify more detailed evolution studies from the PMS to the MS.
        
\section{Results} \label{sec_results}

Here we present the results of running our tidal evolution model for several combinations of input parameters, which include the mass and metallicity of a star, as well as the mass and initial semi-major axis of the companion (a massive planet).
\begin{figure*}[h!]   
\centering
\includegraphics[width=180mm]{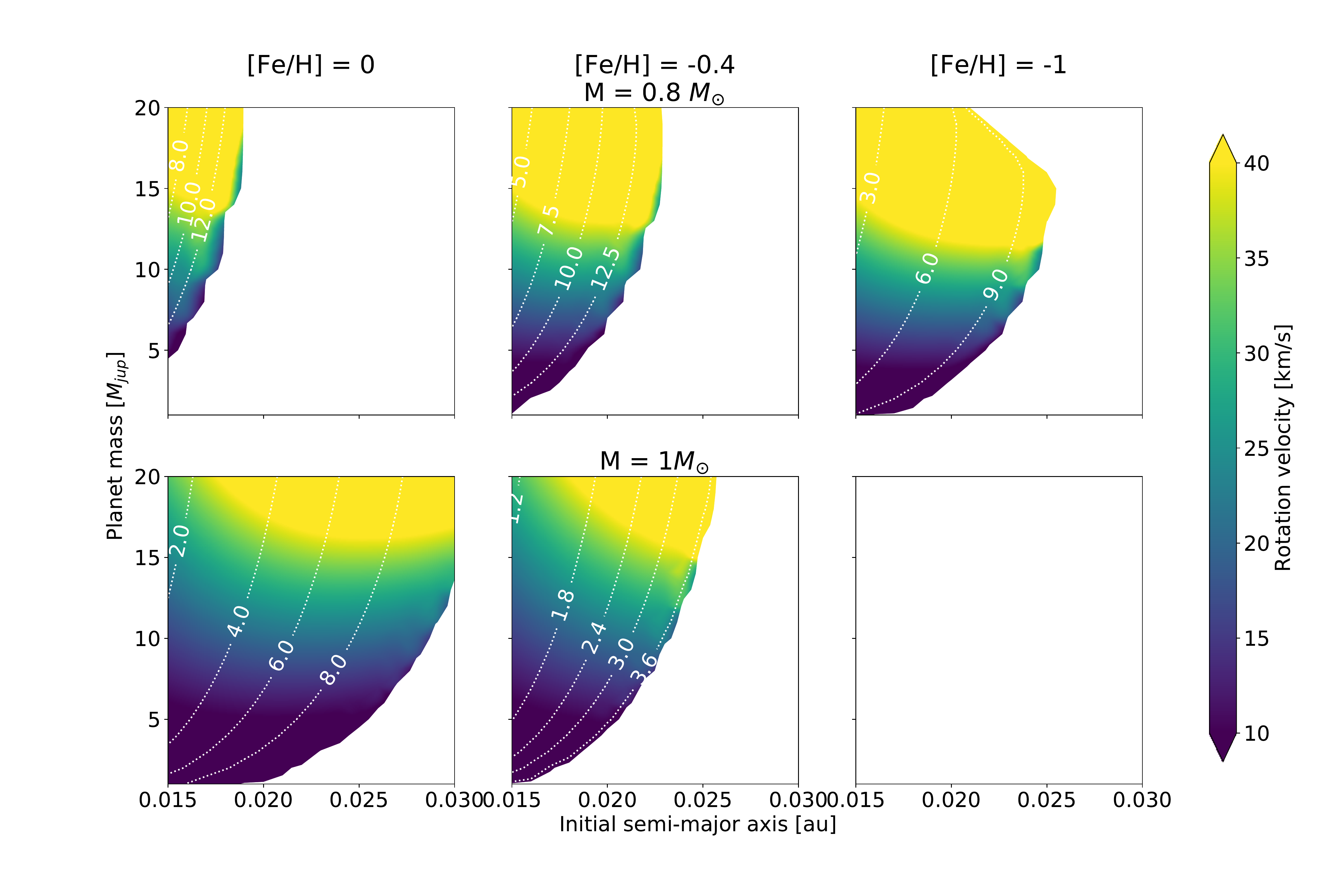}
\caption{Tidal maps: Computational results. Colour-coded is the stellar spin-up depending on the planet mass and initial semi-major axis. The white lines give the time of engulfment in Gyr. The white areas correspond to regions where no engulfment occurs.}
\label{fig_maps}
\end{figure*}
Figure \ref{fig_maps} shows the rotation rate of the star after engulfment in the $a_0 - M_{\rm pl}$ plane, that is, the initial semi-major axis of the planet (x-axis) and its mass (y-axis). We show six stellar models, for different metallicities ([Fe/H] $= 0, -0.4, -1$) and masses (M $= 0.8, 1\,\msun$). The age of the star at the time of engulfment is also indicated. The simulation stops either when the companion reaches the Roche zone, which causes its tidal disruption, when the star reaches the MS turn off before planetary engulfment has occurred, or when the mass of the convective envelope drops below the critical limit. The latter two situations are represented by blank spaces in the upper and lower right corner of the graph. Systems in double synchronicity (Sect. 2.3) are not explored and are represented as blank spaces in the upper right corner of the tidal maps. Clearly, $a_0$ and $M_{\rm pl}$ are two important parameters that influence the outcome of the simulation: more massive planets and planets on larger orbits lead to a more efficient spin-up of a star (see also Fig. 5), because most of the angular momentum is stored in the orbit of the planet around the star. On the other hand, the time to engulfment increases with increasing the initial semi-major axis of the planet.

Comparing the results for the $0.8$ and $1\,\msun$ models, we find that engulfment is much more efficient - and thus, more likely - for more massive stars: the $1\,\msun$ star (Fig. \ref{fig_maps}, bottom row) experiences a significant spin-up within the entire parameter space explored in this work. This, however, is only true as long as a convective envelope thick enough for efficient tidal interaction ($M_{\rm env}/M_{\star} \geq 0.005$)  is present. Therefore, at [Fe/H] = $-1$, the $1\, \msun$ is excluded as the mass of the convective envelope is already below this limit. Even low-mass companions, $M_{\rm pl} < 10\,\mjup$, on orbits from $0.015$ to $0.03$ AU are capable of transferring enough angular momentum to spin-up the star by up to $20$ kms$^{-1}$ from the initial rotation rate of 2 kms$^{-1}$. Also, the engulfment timescales are much shorter for more massive stars. For the $1 \, \msun$ star at [Fe/H] = $-0.4$, the short-period massive planet will be engulfed within $1$ to $3$ Gyr, whereas for the $0.8\,\msun$ star the process will take up to more than 12 Gyr. On the other hand, comparing the amplitudes of spin-up (i.e. the resulting rotation velocity of a star), we find that the impact on the less-massive star is larger. For example, the engulfment of a planet with $M_{\rm pl} \approx 15\,\mjup$ and the initial semi-major axis of $0.015$ AU increases the rotation rate of the $0.8\,\msun$ star by a factor of 40. In contrast, the more massive star with a similar companion will only spin-up by a factor of 25, although the engulfment occurs much faster. This is because, efficient tidal interaction requires a thick convective envelope to dissipate tidal energy within the star. Adding to that is the effect that smaller mass stars have lower moments of inertia because of smaller mass and smaller radii on the MS.

Also the metallicity of the star has a significant influence on the outcome of the star-planet interaction. This is best seen by comparing our results for $\hbox{[Fe/H]}=0$~and $-1$ model (Fig. \ref{fig_maps}, top panel). A solar metallicity $0.8\,\msun$ star requires more than $12$ Gyr to engulf a $10\,\mjup$ companion at an initial distance of $0.017$ AU. This results in a rotation velocity of $\sim$ 30 kms$^{-1}$. This process is three times faster for a star of the same mass, but much lower metallicity, [Fe/H] $=-1$: The companion is engulfed in less than $5$ Gyr. The chemical composition of a star is therefore a very important parameter, in addition to the mass of a star, that has a strong influence on the evolution of the orbit of the planet and the likelihood of engulfment.

The behaviour of spin-up rates with stellar mass and metallicity can be explained by the differences in the stellar structure. For a given mass, more metal-poor stars are larger, have smaller convective envelopes, and evolve faster on the MS. Calculations with our model suggest that such a physical configuration - that is, a smaller convective envelope and larger stellar radius - favours a faster engulfment. However, it should be stressed that the convective envelope must be thick enough to efficiently dissipate tidal energy. Also, beyond some critical combination of mass and metallicity (here $1\,\msun$ and [Fe/H] $= -1$), planetary engulfment is no longer possible, as the mass of the convective envelope is too low for efficient tidal dissipation \citep{Winn2010}. 
\section{Discussion} \label{sec_discussion}
Our results suggest that tidal interaction between a star and its companion -  a massive planet or a brown dwarf - may lead to a significant spin-up of the host star. This can be the consequence of the engulfment of a companion, but it is also true in the case of the double synchronous rotation - a stable configuration that prevents the engulfment and disruption of a companion. The timescales of interaction can vary significantly, from a few hundred Myr up to $\gtrsim$ 12 Gyr, and they depend on several parameters, among them the initial semi-major axis of the planet, the mass of the planet and its host star, and stellar metallicity.

The main prediction of our model is that at low metallicities a larger fraction of planets in the conditions considered in this study are engulfed by their host star, resulting in abnormal rotation velocities. Indeed, lower-mass stars have significantly longer MS lifetimes, and lower metallicity implies a smaller convective envelope, shorter convective turnover time and larger stellar radius, thus favouring a faster engulfment. While the impact of an increasing stellar radius is obvious from the evolution of the semi-major axis, the rate of orbital decay is also a function of the mass of the convective envelope and the convective turnover time ($\tau$). Since the numerical factor is $f \sim 1/\tau^2$ in all our scenarios, the rate of orbital decay follows as: 

\begin{equation}
    \left(\frac{\dot{a}}{a}\right)_{tidal} \sim \left(\frac{M_{env}}{\tau^3} \right)
\end{equation}

With decreasing metallicity it is $M_{env}/\tau^3 > 1$, increasing the orbital decay rate and thus leading to faster planetary engulfment. This was also shown by \citet{Amard2020}, who illustrate changes in the convective turnover time for metal-poor ([Fe/H]=-1) stars.

The prediction of abnormally fast rotation velocities, that surpass any model predictions, in particular for stars with sub-solar metallicity, is a strong observational signature. It could, in principle, be detectable in samples of stars with accurate measurements of ages, metallicities, and rotation periods. It depends, however, on the occurrence rate of close giant planets across stellar metallicities. We have, therefore, analysed multiple catalogues in the literature in the attempt to find stars with reliable estimates of metallicity and short rotation periods \citep{huber, mcquillan}. However, unfortunately, it turned out the current sample statistics is too small and does not allow us to probe the critical metallicity range, [Fe/H]$\lesssim -1$. Most likely, this is because metallicity estimates of fast rotators are difficult, as the spectral lines are significantly broadened and can barely be distinguished from the continuum at [Fe/H]$\lesssim -1$, especially if the signal-to-noise ratio of the observed data is not very high. This implies that targeted surveys must be performed to find fast-rotators, which are candidates for being old and metal-poor.

The lack of a significant population of observed stars with abnormal rotation is not an evidence for the failure of our model. Complementary insights on the feasibility of this scenario can be gained from the observations of planets in discs. A few planetary companions and planetesimals on short orbital periods are known that are currently spiralling inwards toward their host stars. \citet{Vanderburg2015} discovered an evaporating small gas planet around the white dwarf WD 1145+017. \citet{Gaensicke2019} reported a disintegrating dust ring around the white dwarf WD J091405.30+191412.25. WASP-19b \citep{Espinoza2019} and NGTS-10b \citep{McCormac2020} are hot Jupiters on ultra-short orbital periods ($<1$~days); they may spiral towards their host star within a few tens of Myr, leading to a significant spin-up of the host star.

Another interesting consequence of our planetary engulfment scenario is that it could influence the surface chemical composition of the star. Indeed, peculiar abundances were reported for different types of stars in single and binary systems, for example, solar twins, FG-type MS binaries, and white dwarfs \citep[e.g.][]{melendezbergemann,Spina2015,Teske2016,Petrovich2017,Nissen2017, Veras2019}. A recent study by \citet{Oh2018} provides interesting observational evidence for the accretion of a massive rocky planet ($\sim 15\, M_{earth}$) in a $\sim 4$ Gyr old binary system.  Detailed theoretical predictions for this process do not exist yet, but it has already been shown that even the accretion of dust and gas from a proto-planetary disc may lead to a significant alteration of the surface abundances of stars (\citealt{Kunitomo2018}, Hoppe et al. subm). We leave the analysis of the potential influence of the engulfment on the stellar chemical composition to the next paper in the series.

Finally, we would like to briefly discuss our work in the context of another scenario that was recently proposed by \citet{Amard2020}. They suggest that the  magnetic braking phenomenon is not efficient enough to slow down the rotation rate for metal-poor systems. The modified equations introduced by \citet{Amard2020} are, however, conceptually very similar to the equations that we used to simulate spin down (Section~\ref{sec:rot}). Thus, it will be easy to adapt our model framework to simulate both effects in conjunction and to produce an even clearer picture on if one of the mechanisms (tidal engulfment or inefficient magnetic braking) or both may be responsible for rapidly-rotating low-metallicity stars. 
\section{Conclusions} \label{sec_conclusion}
In this work, we study the influence of tidal interaction between a star and its companion on the rotation velocity of the star. This dynamical interaction is followed using a detailed numerical framework that combines tidal friction, magnetic braking, planet migration, and detailed stellar structure models from the GARSTEC grid. We focus on close-in massive planetary companions with masses from 1 to $20\,\mjup$ orbiting low-mass, $0.8$ and $1\,\msun$, MS stars within a broad metallicity range from [Fe/H] $=-1$ to solar.

We find that planetary engulfment of massive planetary companions on tight orbits around their host stars can significantly accelerate stellar rotation on timescales of several billion years. Companions with masses between $5$ and $20\,\mjup$ (the latter is the maximum mass probed in our simulation) and initial semi-major axis $<0.03$~AU lead to a spin-up of the star up to $40$ kms$^{-1}$. 

All parameters, the initial semi-major axis of the planet and its mass, the mass and metallicity of the star, are relevant in determining the final rotation rate of the star. In most cases, the model predicts that the companion is engulfed and the timescales of the process are shorter for more massive host stars. Tidal migration timescales are also shorter for metal-poor stars. This is because low-metallicity systems have a larger radius and a less massive convective envelope that facilitates efficient tidal interaction and favours engulfment of the companion. This, however, is only true as long as the convective envelope is above the limit of $0.005 \, \msun$. Due to the higher mass and, therefore higher moment of inertia, of a $1\,\msun$ star  compared to its low-mass ($0.8\,\msun$) counterpart, the less massive star reaches higher rotation velocities.
We also find that the fastest planetary engulfment occurs for the model with $1\,\msun$ and [Fe/H] $= -0.4$. The largest stellar rotational velocities, for a given planetary mass and initial semi-major axis, are reached for $0.8\,\msun$ and [Fe/H] $= -1$.

In certain ranges of the investigated parameter space, however, no engulfment occurs and the planet forms a stable synchronized system with its host stars. This outcome is seen, for example, for the $0.8 \, \msun$ and [Fe/H] $= -1$ model. Also in this case, the rotation of the host star is significantly increased. 

Our model predicts that low-mass MS stars, owing to the tidal interaction with their close-in companion, can experience a significant spin-up in their surface rotation. Furthermore, stars with abnormal rotation velocities should be more common at low-metallicity, if occurrence rate of close-in massive planets is similar as for more metal-rich stars, as lower [Fe/H] favours faster planet engulfment. Current observational samples are too small and incomplete and thus, do not allow us to test our model. However, future studies, targeting low-metallicity rapidly-rotating stars, can provide enough statistics to test the main observable signature of our scenario. Also the system in a stable synchronized state should, in principle, be detectable via multi-epoch radial velocity measurements. 

In any case, it is clear that for the study of tidal interactions of close-in planets (in our model a$_0 \leq$ 0.03~AU), the evolution of stellar rotation is an important parameter. This is true for the PMS \citep{Rao2018,Bolmont2017} and for the MS stars. The investigation of abnormal fast rotation in metal-poor MS stars is further immensely valuable to investigate several fundamental processes: to probe the limits of applicability of magnetic braking to explain the stellar rotation evolution of MS stars, and further, investigate massive planet and low-mass brown dwarf formation in close vicinity of metal-poor solar-like stars. 

\begin{acknowledgements}
      L.C. acknowledges support by the DFG grant~CA 1795/3. We acknowledge support by the Collaborative Research centre SFB 881 (projects A5, A10), Heidelberg University, of the Deutsche Forschungsgemeinschaft (DFG, German Research Foundation). A.S. is partially supported by MICINN grants ESP2017-82674-R and PID2019-108709GB-I00, and grant 2017-SGR-1131 from Generalitat of Catalunya. We thank the anonymous referee for the insightful and constructive comments.
\end{acknowledgements}

 \bibliographystyle{style} % style style.bst
 \bibliography{ref} % your references Yourfile.bib

\end{document}